\documentclass[12pt,a4]{article}
\usepackage{latexsym}

\title{Normal Modes and No Zero Mode Theorem 
of Scalar Fields in BTZ Black Hole Spacetime}
\author{Masakatsu KENMOKU \thanks{kenmoku@asuka.phys.nara-wu.ac.jp}, \ 
Maiko KUWATA \thanks{kuwata@asuka.phys.nara-wu.ac.jp} \\
Department of Physics, Nara Women's University, Nara 630-8506, Japan
\\and \\
Kazuyasu SHIGEMOTO \\
Tezukayama University, Nara 631-8501, Japan
\thanks{shigemot@tezukayama-u.ac.jp}}
\date{\empty}
\begin{document}
\maketitle
\abstract{
Eigenfunctions for normal modes of 
scalar fields in BTZ black hole spacetime are studied. 
Orthonormal relations among them are derived.  
Quantization for scalar fields is done and 
particle number, energy and angular momentum 
are expressed by the creation and annihilation operators.   
Allowed physical normal mode region is studied 
on the basis of the no zero mode theorem. 
Its implication to the statistical mechanics is also studied.    
}

\section{Introduction}
One of the important aspects 
of general relativity is black hole physics.
Extensive studies in this field have been accomplished 
from the observational and/or theoretical 
view points.
Many evidences of black holes have been observed   
including the supermassive black holes 
at the center of the galaxies \cite{eckart001,herrnstein001}.  
Such black holes are expected to be well described 
by axisymmetric solutions of the Einstein's field equation. 
Recently multi-parameter rotating black hole solutions 
in higher dimensions are studied theoretically 
from the view point of string theory, M theory, brane world,  
and with the interest of AdS/CFT correspondence 
\cite{myers001,gibbons001,chen001}. 

One of the most interesting aspects of black hole physics 
is the thermodynamical properties. 
Black holes can be considered as thermal objects 
\cite{bekenstein001,bardeen001}: 
the entropy is  
proportional to surface area on the horizon 
with the Hawking temperature $T_{\rm H}$ 
\cite{hawking001}. 
The statistical understanding to the thermodynamics of black holes 
has been tried in the frames of field theories 
\cite{birrell001}, string theory  
\cite{strominger001} and others. 
Among them, the brick wall model has been proposed by 
't Hoof in order to interpret the area law of the black hole entropy 
by considering the freedom of scalar fields 
around black hole horizon in the standard framework 
of statistical mechanics \cite{thooft001}.        
  
On the other hand, there are some problematic issues 
in understanding the black hole thermoynamics. 
One of fundamental problems is the super-radiant instability, 
which occurs in case of rotating black holes.  
This problem is the case that 
the flux intensity of scattered outgoing fields to black hole 
becomes larger than that of ingoing fields 
under the condition of $\omega-\Omega_{\rm H}m < 0$ 
\cite{bardeen002,cardoro001}, 
where $\omega$ and $m$ are frequency and azimuthal angular momentum 
of field and $\Omega_{\rm H}$ is the angular velocity of black 
hole.  
The Boltzmann factor and then the partition function become 
ill-defined due to the existence of the super-radiance 
\cite{mukohyama001}. 
In BTZ black hole spacetime \cite{btz001}, 
the super-radiance problem for the scalar fields 
is also discussed extensively 
\cite{ichinose001,swkim001,fatibene001,ho001}. 
In our previous papers, statistical mechanics 
of scalar fields in multi-rotating black hole spacetime 
are studied \cite{kenmoku001,kenmoku002}. 

The purpose of this paper is to investigate 
the normal modes of scalar fields around rotation black holes 
to define the statistical mechanics well and to 
understand the super-radiance problem. 
In order to make the problem clear, 
we study the BTZ black hole model which is 
the anti-de Sitter rotating black hole in (2+1) dimension. 
In case of quasinormal modes in BTZ spacetime, 
exact analytically treatment have been done \cite{birmingham001}. 
Therefore we expect to obtain the exact eigenfunctions 
in case of normal modes too.     
As the boundary conditions, 
we impose the Dirichlet boundary condition at infinity and 
the Dirichlet or Neumann boundary condition at horizon 
for eigenfunctions of normal modes.

The organization of this paper is as follows. 
In section 2, we prepare notations and definitions  
of BTZ black hole spacetime and scalar fields  
for the convenience of the subsequent sections.   
In section 3, we derive the orthonormal relations among 
eigenfunctions of normal modes. 
In section 4, 
the quantization of the scalar fields will be done. 
In section 5, particle number, energy and angular momentum  
are calculated to represent by creation and annihilation operators.  
In section 6, we will show the theorem of 
nonexistence of zero normal mode eigenstates.  
The allowed physical normal mode region 
will be derived as its application   
and statistical implication will also be studied in the section too.  
The results will be summarized in the final section.

\section{Scalar fields in BTZ black hole spacetime}
\renewcommand{\theequation}{\thesection.\arabic{equation}}
\setcounter{equation}{0}
This section is the preparation for definitions and notations 
in the following sections.

The Einstein-Hilbert action with negative cosmological constant 
($\Lambda=-1/\ell^2$) in (2+1) dimension is 
\begin{eqnarray}
I_{G}=\frac{1}{16\pi G}\int d^3x \sqrt{-g}(R+\frac{2}{\ell^2})\ .
\end{eqnarray}
In the following, the gravitational constant is normalized as 
$G=1/8$D 
The vacuum Einstein's equations for this action are 
\begin{eqnarray}
R_{\mu\nu}-\frac{g_{\mu\nu}}{2}R-\frac{g_{\mu\nu}}{\ell^2}=0\ .
\end{eqnarray}
Banados, Teitelboim and Zanelli (BTZ) solution is in the form as 
\cite{btz001}, 
\begin{eqnarray}
ds^2=g_{tt}dt^2+g_{\phi \phi}d\phi^2+2g_{t\phi}dtd\phi+g_{rr}dr^2 \ ,
\end{eqnarray}
where the components of metric are 
\begin{eqnarray}
g_{tt}&=&M-\frac{r^2}{\ell^2}\ , \ \ \  g_{t\phi}=-\frac{J}{2}\ , \\
g_{\phi\phi}&=&r^2\ , 
\ \ \ g_{rr}=(-M+\frac{J^2}{4r^2}+\frac{r^2}{\ell^2})^{-1} 
\end{eqnarray}
with black hole mass $M$ and rotation parameter $J$. 
The contravariant time component of metric is 
negative of covariant radial component:  
\begin{eqnarray}
g^{tt}=-g_{rr}\ .
\end{eqnarray}
Zeros of their inverse function denote the horizon of black hole  
\begin{eqnarray}
r_{\pm}^2=\frac{M\ell^2}{2}\left({1\pm\sqrt{1-\frac{J^2}{M^2\ell^2}}}\right)
\ , 
\end{eqnarray}
where the event horizon is $r_{+}$.  
Note that the BTZ metrics can be rewritten in a diagonal form: 
\begin{eqnarray}
ds^2=\frac{1}{g^{tt}}dt^2
+g_{\phi\phi}(d\phi+\frac{g_{t\phi}}{g_{\phi\phi}}dt)^2
+g_{rr}dr^2\ .
\end{eqnarray}

The action and the Lagrangian density for the minimally coupled 
complex scalar field with dimensionless mass $\mu$ 
under the BTZ spacetime is 
\begin{eqnarray}
I_{\rm Scalar}&=& \int dt dr d\phi \sqrt{-g} \, {L}_{\rm Scalar} \ , \\ 
{L}_{\rm Scalar}&=&
-(g^{\mu\nu} \partial_{\mu}\Phi^* (x)\partial_{\nu}\Phi(x)
+\frac{\mu}{\ell^2}\Phi^*(x)\Phi(x))\ .
\end{eqnarray}
The Klein - Gordon equation for this action is 
\begin{eqnarray}
	\left(\frac{1}{\sqrt{-g}}
\partial_{\mu}(g^{\mu\nu}\sqrt{-g}\partial_{\nu})
 - \frac{\mu}{l^2} \right)\Phi = 0 \ . \\
\end{eqnarray}
The background BTZ metric do not depend on time and azimuthal 
angle variables,  
and then the scalar field solution is put in the form: 
\begin{eqnarray}
	\Phi 
={\rm e}^{-i\omega t}{\rm e}^{i m \phi} R(r) \ , 
\end{eqnarray}
where $\omega$ and $m\ (m=0,\pm1,\pm2,\cdots)$ 
denote frequency and azimuthal angular momentum. 
The equation for the radial wave function becomes 
\begin{eqnarray}
\left(- g^{tt}(\omega-\frac{J}{2r^2}m)^2-\frac{m^2}{r^2}
+\frac{1}{r}\frac{d}{dr}\frac{r}{g_{rr}}
\frac{d}{dr}-\frac{\mu}{\ell^2} \right)R(r)=0\ .
\end{eqnarray}
We note that this equation is invariant under 
the symmetry of $(\omega,m)$ and $(-\omega,-m)$. 

\section{Normal modes and orthonormal relations }
\setcounter{equation}{0}
\noindent
We consider the radial wave equation as the eigenvalue equation 
for the eigenfunction $R_{\omega,m } $
with eigenvalues $\omega$ and $m$:   
\begin{eqnarray}
\Delta_{r}R(r)_{\omega,m}&=&\left(g^{tt}(\omega-\Omega_{r}m)^2
+\frac{m^2}{r^2}+\frac{\mu}{\ell^2}\right)R(r)_{\omega,m} \ , 
\end{eqnarray}
where Laplacian and the angular velocity at radial position $r$ 
are defined by:
\begin{eqnarray} 
\Delta_{r}&:=& 
\frac{1}{r}\frac{d}{dr}\frac{r}{g_{rr}}\frac{d}{dr}\ , \nonumber\\  
\Omega_{r}&:=&-\frac{g_{t\phi}}{g_{\phi\phi}}
=\frac{g^{t\phi}}{g^{tt}}
=\frac{J}{2r^2}\ .
\end{eqnarray}
The following two integrations are considered 
\begin{eqnarray}
&&\int dr \sqrt{-g}R_{\omega,m}^{*}\Delta_{r}R_{\omega',m}\nonumber\\
&&=\int dr \sqrt{-g} \left(g^{tt}(\omega'-\Omega_{r}m)^2
+\frac{m^2}{r^2}+\frac{\mu}{\ell^2}\right)
R_{\omega,m}^{*}R_{\omega',m}\ , \\
&&\int dr \sqrt{-g}\Delta_{r}R_{\omega,m}^{*}R_{\omega',m}\nonumber\\
&&=\int dr \sqrt{-g} \left(g^{tt}(\omega-\Omega_{r}m)^{2}
+\frac{m^2}{r^2}+\frac{\mu}{\ell^2}\right)^{*}
R_{\omega,m}^{*}R_{\omega',m}\ .
\end{eqnarray}
In cases of vanishing the boundary terms, 
the difference of these integrations becomes    
\begin{eqnarray}
0&=&
\int dr \sqrt{-g} 
\left(-
\left(g^{tt}(\omega-\Omega_{r}m)^{2}
+\frac{m^2}{r^2}+\frac{\mu}{\ell^2}\right)^{*}
+
\left(g^{tt}(\omega'-\Omega_{r}m)^{2}
+\frac{m^2}{r^2}+\frac{\mu}{\ell^2}\right)\right)\nonumber\\
&\times&R_{\omega,m}^{*}R_{\omega',m}\nonumber \\
&=&
\int dr \sqrt{-g}\,g_{rr}\, 
(\omega^{*}-\omega')(\omega^{*}+\omega'-2\Omega_{r}m)
R_{\omega,m}^{*}R_{\omega',m}\ .
\end{eqnarray}
For the case of $\omega=\omega'$, 
this relation shows the reality of eigenvalue $\omega$. 
For the case of $\omega \neq \omega'$, 
this relation shows the orthonormal relations among eigenfunctions: 
\begin{eqnarray}
\int dr \sqrt{-g}\,(-g^{tt})\, 
(\omega+\omega'-2\Omega_{r}m)
R_{\omega,m}^{*}R_{\omega',m}\ =\delta_{{\omega}, \omega'}\ , 
\end{eqnarray}
Similarly, we have orthogonal relation 
for a couple of positive and negative azimuthal angular momentum 
$(m,-m)$ as
\begin{eqnarray}
\int dr \sqrt{-g}\,g^{tt}\, 
(\omega-\omega'-2\Omega_{r}m)
R_{\omega,m}R_{\omega',-m}\ =0\ .
\end{eqnarray}

Defining the full eigenfunction with normalization factor 
$N_{\omega,m}$ 
\begin{eqnarray}
f_{\omega,m}:=N_{\omega,m}{\rm e}^{-i\omega t}{\rm e}^{im\phi}R_{\omega,m}\ , 
\end{eqnarray}
we have the full orthonormal relations: 
\begin{eqnarray}
\int_{\Sigma} d\phi dr \sqrt{-g}\, (-g^{tt}) \, 
(\omega+\omega'-2\Omega_{r}m)\,
f_{\omega,m}^{*}f_{\omega',m'}\ 
&=&\delta_{\omega,\omega'}\delta_{m,m'}\ , \nonumber\\
 \int_{\Sigma} d\phi dr \sqrt{-g}\, (-g^{tt}) \, 
(\omega-\omega'-2\Omega_{r}m)\,
f_{\omega,m}f_{\omega',m'}\ 
&=&0 \ , 
\end{eqnarray}
where the integration region $\Sigma$ denotes $0\leq\phi<2\pi$ 
and $r_{\rm H}\leq r <\infty$ 
\footnote{
In order to regularize the divergence on the horizon,  
the cutoff parameter is introduced in explicit construction 
of normal mode solutions \cite{kuwata001}.  
The cutoff parameter plays the similar role 
in the brick wall model \cite{thooft001} }. 
For more compact notation, the inner products are introduced:  
\begin{eqnarray}
(A,B):=
\int_{\Sigma}d\phi dr \sqrt{-g}(-ig^{t\nu})
\left(A^{*}(x)\partial_{\nu}B(x)
-\partial_{\nu}A^{*}(x)B(x)\right)\ ,
\end{eqnarray}
and the general form of orthonormal relations are obtained:  
\begin{eqnarray}
(f_{\omega,m},f_{\omega',m'})&=&-(f_{\omega,m}^{*},f_{\omega',m'}^{*})
=\delta_{\omega,\omega'}\delta_{m,m'}\nonumber \\
(f_{\omega,m}^{*},f_{\omega',m'})&=&(f_{\omega,m},f_{\omega',m'}^{*})
=0 \ .
\end{eqnarray}
It should be noted that the orthonormal relations hold for 
the allowed normal mode region $0 < \omega-\Omega_{r}m$.

\section{Quantization }
\setcounter{equation}{0}
\noindent
We can define the canonical momentum conjugate to scalar field $\Phi$ 
as 
\begin{eqnarray}
\Pi:=\frac{\partial {L}_{\rm Scalar}}{\partial \, \partial_{t}{\Phi}}
=-g^{t\nu}\partial_{\nu}\Phi^{\dagger}
=-g^{tt}(\partial_{t}{\Phi^{\dagger}}+\Omega_{r}\partial_{\phi}\Phi^{\dagger})
\ , 
\end{eqnarray}
where $\dagger$ denotes the Hermitian conjugate operation 
for the quantized fields.
We impose the equal time commutation relations  
among fields and their momenta:  
\begin{eqnarray}
\left[\Phi(t,r, \phi),\Pi(t,r', \phi')\right]
&=&\frac{i}{\sqrt{-g}}\delta(r-r') \delta(\phi-\phi')\ , \nonumber\\
\left[\Phi(t,r, \phi)^{\dagger},\Pi(t,r', \phi')^{\dagger}\right]
&=&\frac{i}{\sqrt{-g}}\delta(r-r') \delta(\phi-\phi')\ ,  
\end{eqnarray}
and others are zero.
We make the normal mode expansion for 
the scalar fields and the conjugate momenta as 
\begin{eqnarray}
\Phi&=&\sum_{\omega,m}\left(
a_{\omega,m}f_{\omega,m}+b_{\omega,m}^{\dagger}f_{\omega,m}^{*}
\right)
\ , \nonumber \\
\Pi&=&-ig^{tt}
\sum_{\omega,m}(\omega-\Omega_{r}m)
(a_{\omega,m}^{\dagger}f_{\omega,m}^{*}-b_{\omega,m}f_{\omega,m})
\ .
\end{eqnarray}
Expansion coefficients are inversely expressed 
by fields and their momenta using the orthonormal relations: 
\begin{eqnarray}
a_{\omega,m}
&=&(f_{\omega,m},\Phi) \nonumber\\
&=& \int_{\Sigma} d\phi dr \sqrt{-g} \left(
i f_{t,\omega,m}^{*}(t,r,\phi)\Pi^{\dagger}(t,r,\phi) \right. \nonumber\\
&&\left. 
-g^{tt}(\omega-\Omega_{r}m)f_{t,\omega,m}^{*}(t,r,\phi)\Phi(t,r,\phi)
\right) \ ,\nonumber\\
b_{\omega,m}^{\dagger}
&=&-(f_{\omega,m}^{*},\Phi)\nonumber\\
&=&- \int_{\Sigma} d\phi dr \sqrt{-g} \left(
i f_{t,\omega,m}(t,r,\phi)\Pi(t,r,\phi)^{\dagger} \right. \nonumber\\
&&\left. 
+g^{tt}
(\omega-\Omega_{r}m)f_{t,\omega,m}(t,r,\phi)\Phi(t,r,\phi)
\right)\ .
\end{eqnarray}
Completeness relations are derived from the 
consistency of normal mode expansions of fields: 
\begin{eqnarray}
&&\sum_{\omega,m}(-g^{tt})(\omega-\Omega_{r}m)
\left(f_{\omega,m}(t,r,\phi)f_{\omega,m}(t,r',\phi')^{*}
+f_{\omega,m}(t,r,\phi)^{*}f_{\omega,m}(t,r',\phi')\right)\nonumber \\
&=&\frac{1}{\sqrt{-g}}\delta(r-r')\delta(\phi-\phi')\ , \nonumber \\
&&\sum_{\omega,m}
\left(f_{\omega,m}(t,r,\phi)f_{\omega,m}(t,r',\phi')^{*}
-f_{\omega,m}(t,r,\phi)^{*}f_{\omega,m}(t,r',\phi')\right)=0\ .
\end{eqnarray}
The commutation relations between annihilation and creation operators 
are derived using the completeness relations: 
\begin{eqnarray}
[a_{\omega ,m},a_{\omega ',m'}^{\dagger}]
=\delta_{\omega ,\omega '}\delta_{m,m'}\ ,  
[b_{\omega ,m},b_{\omega ',m'}^{\dagger}]
=\delta_{\omega ,\omega '}\delta_{m,m'}\ , 
\end{eqnarray} 
and others are zero.
  
\section{Particle number, energy and angular momentum}
\setcounter{equation}{0}
\noindent
 
>From the symmetry of the action for the scalar filed, 
conserved particle number, energy and angular momentum 
are derived and expressed by the creation and annihilation operators.  
\begin{itemize}
\item[1.] Particle number \\
>From the phase translation invariance of the action, 
particle number current is defined  
\begin{eqnarray}
j^{\mu}&=&-i g^{\mu\nu}
(\Phi^{\dagger}\partial_{\nu}\Phi-\partial_{\nu}\Phi^{\dagger}\Phi)\ ,
\end{eqnarray}
which is shown to satisfy the current conservation  
\begin{eqnarray}
j^{\mu}_{\ ;\mu}=0 \ .
\end{eqnarray}
The corresponding particle number is 
\begin{eqnarray}
N:&=& \int_{\Sigma} d{\phi} dr \sqrt{-g}\, j^{\,t} \nonumber\\
  &=& \int_{\Sigma} d\phi dr \sqrt{-g}\,
(\Pi\Phi-\Phi^{\dagger}\Pi^{\dagger})\ . 
\end{eqnarray}
The particle number is also expressed by the creation and annihilation 
operators by using normal mode expansion of fields 
and their momenta 
\begin{eqnarray}
N=\sum_{\omega,m}(a_{\omega,m}^{\dagger}a_{\omega,m}
-b_{\omega,m}b_{\omega,m}^{\dagger})\ .
\end{eqnarray}
\item[2.] Energy and Angular Momentum \\
Because the metrics don't depend on time and azimuthal angle, 
two Killing vectors exist 
\begin{eqnarray}
\xi_{(t)}^{\mu}=(1,0,0)\ \ \ , \ \ \ \xi_{(\phi)}^{\mu}=(0,0,1)
\ .
\end{eqnarray}
Defining the energy-momentum tensor  
\begin{eqnarray}
T_{\mu\nu}&=&-\frac{2}{\sqrt{-g}}\frac{\delta I_{\rm M}}{\delta g_{\mu\nu}}
\nonumber\\
&=&\partial_{\mu}\Phi^{\dagger}\partial_{\nu}\Phi
+\partial_{\nu}\Phi^{\dagger}\partial_{\mu}\Phi
-g_{\mu\nu}(g^{\alpha\beta}\partial_{\alpha}\Phi^{\dagger}\partial_{\beta}\Phi
+\frac{\mu}{\ell^2}\Phi^{\dagger}\Phi)\ , 
\end{eqnarray}
local conservation laws hold for two Killing vectors 
\begin{eqnarray}
(\xi_{(i)}^{\mu} T_{\mu}^{\nu})_{;\nu}=0 \ \ , \ \ 
{\mbox for}\ \ i=t, \phi \ .
\end{eqnarray}
Corresponding conservative quantities are 
energy and angular momentum   
\begin{eqnarray}
E&=&- \int_{\Sigma} 
d\phi dr \sqrt{-g}\,(\xi_{(t)}^{\mu}T_{\mu}^{t})\nonumber\\
&=& \int_{\Sigma} d\phi dr\sqrt{-g}(-g^{tt}\Phi^{\dagger}_{,t}\Phi_{,t}
+g^{\phi\phi}\Phi^{\dagger}_{,\phi}\Phi_{,\phi}
+g^{rr}\Phi^{\dagger}_{,r}\Phi_{,r})\\
L&=& \int_{\Sigma} 
d\phi dr\sqrt{-g}\,(\xi_{(\phi)}^{\mu}T_{\mu}^{t})\nonumber\\
&=& \int_{\Sigma} d\phi dr
\sqrt{-g}(g^{tt}(\Phi^{\dagger}_{,t}\Phi_{,\phi}
+\Phi^{\dagger}_{,\phi}\Phi_{,t})
+2g^{t\phi}\Phi^{\dagger}_{,\phi}\Phi_{,\phi}) \ .
\end{eqnarray}
The energy and angular momentum are
expressed by the creation and annihilation 
operators by using normal mode expansion of fields 
and their momenta 
\begin{eqnarray}
E&=&\sum_{\omega,m}\omega(a_{\omega,m}^{\dagger}a_{\omega,m}
+b_{\omega,m}b_{\omega,m}^{\dagger})\ , \\
L&=&\sum_{\omega,m}m(a_{\omega,m}^{\dagger}a_{\omega,m}
+b_{\omega,m}b_{\omega,m}^{\dagger})\ . 
\end{eqnarray}
The effective energy, which is the energy taking into the rotation effect 
on the horizon, is expressed as 
\begin{eqnarray}
E-\Omega_{\rm H}L
&=& \sum_{\omega,m}(\omega-\Omega_{\rm H}m)(a_{\omega,m}^{\dagger}a_{\omega,m}
+b_{\omega,m}b_{\omega,m}^{\dagger})\ , 
\label{eqn:5012}
\end{eqnarray}
where $\Omega_{\rm H}=J/2r_{+}^2$ is the angular velocity 
on the horizon. The effective energy  
is positive definite for the allowed normal mode region 
$0<\omega-\Omega_{\rm H}m$ except for the zero point energy. 

\section{No zero mode theorem}
\setcounter{equation}{0}
\noindent
In this section, 
we consider the zero mode eigenstates, 
which are defined as the stats of $0=\omega-\Omega_{\rm H}m$ 
with $-\infty <m<\infty$. 
We will show that they do not exist and 
the allowed normal mode region is $0< \omega-\Omega_{\rm H}m$. 
We also show that the statistical mechanics for the 
Hartle-Hawking state is defined well for BTZ black hole spacetime. 
We impose Dirichlet or Neumann boundary condition 
at horizon and Dirichlet boundary condition 
at infinity because spacetime is anti-de Sitter.  
\begin{description}
\item[Statement 1.] Eigenfunction 
of normal mode for $\omega=m=0$ does not exist.
\item[Proof] 
General radial eigenfunction of normal mode 
for $\omega=m=0$ is obtained solving the 
radial wave equation in case of massless scalar fields $\mu=0$ as 
\begin{eqnarray}
R_{o,o}=c_{1}\ln{(\frac{r^2-r_{+}^2}{r^2-r_{-}^{2}})}+c_{2} \ , 
\label{eqn:6001}
\end{eqnarray}
where $c_{1},c_{2}$ are integration constants. 
This solution cannot satisfy both of the boundary conditions 
at horizon and infinity.    
\item[Statement 2. (No zero mode theorem)]
Eigenfunctions of normal modes for $0=\omega-\Omega_{\rm H}m$ 
don't exist.
\item[Proof]
The radial eigenfunctions of normal mode for $0=\omega-\Omega_{\rm H}m$ 
satisfying the boundary condition at infinity 
is obtained by the hypergeometric function as 
\begin{eqnarray}
R_{\omega(=\Omega_{\rm H}m),m}
=(\frac{r_{+}^2-r_{-}^2}{r^2-r_{-}^{2}})^{b}
\frac{1}{\Gamma(2b)}
F(b-ic, b+ic,2b;\frac{r_{+}^2-r_{-}^2}{r^2-r_{-}^{2}})\ ,
\end{eqnarray}
where parameters are 
\begin{eqnarray}
b=\frac{1+\sqrt{1+\mu}}{2}\ \ , \ \ c=\frac{\ell m}{2r_{+}}\ .
\end{eqnarray}
This solution also cannot satisfy the boundary condition  
at horizon. Note that the solution reduces the first term in 
eq. (\ref{eqn:6001}) for $\omega=m=\mu=0$.  
\item[Stament 3.]
The allowed physical normal mode region is 
$0<\omega-\Omega_{\rm H}m$ with $-\infty <m<\infty$.  
\item[Proof]
First remark that 
the allowed physical normal mode region is $0<\omega$ with $-\infty <m<\infty$
for the case of no rotation $J=0$. 
We assume that the normal mode is analytic for rotation parameter $J$. 
After switching on the rotation $J\neq 0$, 
the allowed physical normal mode region shifts from $0<\omega$ to 
 $0<\omega-\Omega_{\rm H}m$ 
because normal modes cannot across each other. 

It should be noted that the mode of $0=\omega-\Omega_{\rm H}m$ 
is a special mode in the sense
that it is the unique solution which 
satisfies the Dirichlet boundary condition at infinity  
but deverges at horizon.    
As a consequence, normal modes are devided into two regions: 
one region  is   
$0 < \omega-\Omega_{\rm H}m$ with $-\infty < m < \infty$ 
and the other region  is  
$0 > \omega-\Omega_{\rm H}m$ with $-\infty < m < \infty$ 
devided by the zero mode. 
We can confirm this result by the explicit construction of 
the normal mode eigenfunctions \cite{kuwata001}. 
 
\item[Statement 4.]
Statistical mechanics for the scalar fields around 
BTZ black hole spacetime is well-defined. 
\item[Proof]
The allowed region for the total effective energy becomes 
$0<E-\Omega_{\rm H}L$ from the expression in eq. (\ref{eqn:5012}).  
Then the Boltzmann factors 
and then the partition function 
taking account of rotating effect by Hartle and Hawking 
\cite{hawking002} 
become well-defined:    
\begin{eqnarray}
Z={\rm Tr}\ {\exp}(-\beta_{\rm H} (E-\Omega_{\rm H}L))\ ,
\end{eqnarray}
where trace is taken for occupation numbers of scalar particles  
and $\beta_{\rm H}$ denotes the inverse of Hawking temperature 
\cite{hawking001}.  
\end{description}

\section{Summary}
We have investigated normal modes of scalar fields around 
BTZ black hole spacetime with the Dirichlet or Neumann boundary 
condition at horizon and convergence to zero at infinity. 
Orthonormal relations and completeness relations 
for normal mode eigenfunctions are derived. 
Using these relations, 
conserved charge, energy and angular momentum are 
expressed by creation and annihilation 
operators of scalar particles. 
Zero modes of $0=\omega-\Omega_{\rm H}m$ 
are shown not to exist and the allowed normal mode region 
is shown to be $0<\omega-\Omega_{\rm H}m$. 
>From the allowed normal mode region, 
the total effective energy should be positive; 
$0<E-\Omega_{\rm H}L$, 
which guarantees the Boltzmann factors 
for the rotating black holes to be well-defined. 
This fact shows that the super-radiance does not 
occur for BTZ black hole spacetime 
and is consistent with the negative value of the imaginary part 
for quasinormal frequency \cite{birmingham001}.

We can extend our result of (2+1) dimensional 
BTZ black hole spacetime to (3+1) or more 
higher dimensional rotating black hole spacetime, 
which will help for the deeper understanding of 
scalar field entropy around the rotating black holes 
\cite{kenmoku003}. 
We also try to study the relation between 
our method and the treatment of the conformal field theory, 
which is developed extensively in view of exact AdS/CFT correspondence 
in BTZ spacetime \cite{cerlip001,ksgupta001,witten001,rkgupta001}.  

\vspace{0.5cm}
\noindent
{{ \Large {\bf Acknowledgements}}}

\vspace{0.3cm}
\noindent
One of authors (M.Kenmoku) would like to thank Professor Gupta Kumar for 
comments about the boundary condition for orthogonality 
of normal modes.

\end{itemize}

\end{document}